# Functional groups accessibility and the origin of photoluminescence in N/O-containing bottom-up carbon nanodots

Paul P. Debes*[a,b], Michal Langer[c], Melanie Pagel[a,b], Enzo Menna[d,e], Bernd Smarsly[a,b], Silvio Osella[c], Jaime Gallego*[a,b], Teresa Gatti*[a,b,f]

| | |
|---|---|
| [a] | P. P. Debes, M. Pagel, Prof. Dr. B. Smarsly, Dr. J. Gallego, Prof. Dr. T. Gatti |
| | Center for Materials Research |
| | Justus-Liebig University |
| | Heinrich-Buff-Ring 17, 35392, Giessen (Germany) |
| | E-mail: paul.debes@phys.chemie.uni-giessen.de ; jaime.gallego-marin@phys.chemie.uni-giessen.de |
| [b] | P. P. Debes, M. Pagel, Prof. Dr. B. Smarsly, Dr. J. Gallego, Prof. Dr. T. Gatti |
| | Institute of Physical Chemistry |
| | Justus-Liebig University |
| | Heinrich-Buff-Ring 17, 35392, Giessen (Germany) |
| [c] | Dr. M. Langer, Prof. Dr. S. Osella |
| | Chemical and Biological Systems Simulation Lab, Centre of New Technologies |
| | University of Warsaw |
| | 2c Banacha Street, 02-097, Warszawa (Poland) |
| [d] | Prof. Dr. E. Menna |
| | Department of Chemical Sciences & INSTM |
| | University of Padova |
| | Via Marzolo 1, 35131, Padova (Italy) |
| [e] | Prof. Dr. E. Menna |
| | Interdepartmental Centre Giorgio Levi Cases for Energy Economics and Technology |
| | University of Padua |
| | Via Marzolo 9, 35131, Padova (Italy) |
| [f] | Prof. Dr. T. Gatti |
| | Department of Applied Science and Technology |
| | Politecnico di Torino |
| | C.so Duca degli Abruzzi 24, 10129, Torino (Italy) |
| | E-mail: teresa.gatti@polito.it |

Supporting information for this article is given via a link at the end of the document.

**Abstract:** Chemical functionalization of carbon nanodots (CNDs) offers a valuable opportunity to tailor multifunctionality in these nanocarbons, by engineering the composition of their molecular surface. Therefore, it is important to elucidate the type and amount of CNDs functionalization to be able to design their properties accurately. CNDs are often functionalized through amide coupling without validating the degree of functionalization. As a measure of functionalization, the amounts of primary amines *via* Kaiser test (KT) or imine reactions of the bare CNDs is often considered. However, this may lead to overestimating the degree of functionalization obtained by the pure amide coupling due to different reaction mechanisms and involved intermediates. Herein, four different CNDs prepared by microwave-assisted synthesis from arginine or citric acid with varying amounts of ethylenediamine are presented. We resorted to a combination of physicochemical methods to provide elemental, structural, and optical information. By that, we developed a method to quantify the degree of functionalization by amide coupling and show that the functionalization is lower than one would expect. Comparing experimental optical features of the CNDs with different computed model systems further allows us to provide a more advanced vision of structure-property relationships in these still elusive nanocarbons.

## Introduction

Due to their valuable electronic, thermal, optical, chemical, and mechanical properties, 0D carbon dots (CDs) have attracted great interest of the chemistry and materials science communities in the last two decades.[1] In more detail, there are three different subclasses of CDs concerning the proportion of $sp^2/sp^3$ carbons,[2,3] namely carbonized polymer dots (CPDs), graphene quantum dots (GQDs), and carbon nanodots (CNDs).[4] All these species are the focus of different research efforts, ranging from purely synthetic/structural to more application-oriented ones. CPDs are obtained by carbonising polymer clusters, and the degree of carbonization plays an important role in forming their polymer/carbon hybrid structure, which defines their properties.[3] In addition to high photoluminescence quantum yields (PLQY), good aqueous solubility, and high oxygen/nitrogen contents, CPDs are unique in having short polymer chains, abundance of functional groups, and highly cross-linked network structures.[5–7] GQDs consist of a graphene/graphite core, and edges that can be modified by functional groups, resulting in good solubility.[8,9] Due to the quantum confinement effect, the fluorescence properties are unrivalled and can be modified for various applications.[10] CNDs are quasi-spherical shaped nanoparticles, and their size is less than 10 nm.[11] Furthermore, they show remarkable PL, solubility in aqueous solutions, and the toxicity has been reported to be low.[12,13] A facile, fast, and inexpensive synthesis is the most significant benefit of CNDs.[14] Different procedures to synthesize CNDs in a bottom-up fashion, starting from variable small molecule-based precursors, are well-studied.[15–17] The bottom-up approach is commonly based on thermal,[18] microwave(MW)-assisted,[19] and hydro-/solvothermal treatments[20] of smaller molecules. Bottom-up approaches make it easier to modulate the amount and variety of functional groups by using a wide range of small molecules as starting materials.[14] Citric acid (CA), for example, resolves in oxygen-containing groups like carboxylic acids, aldehydes, hydroxyls, carbonyls, and epoxides, predominantly.[21–23] While small nitrogen-containing molecules, such as urea, arginine (Arg), or other amino acids, have been used to introduce amine groups into the CNDs.[24,25] One of the reagents used to increase the amine content in the CNDs is also ethylenediamine (EDA).[15,26] Further functionalization of the aforementioned functional groups with small molecules, polymers,[27] ions, metals, DNA, or proteins is used to tune the physicochemical properties of CNDs and their potential applications, including sensing,[28,29] catalysis,[30,31] and optoelectronics.[11,32,33] Due to their photobleaching resistance, chemical inertness, and variety of functional groups, CNDs fulfill the prerequisites for their use in optoelectronic applications.[34] Since the degree of functionalization affects the optoelectronic properties of nanocarbon hybrid materials in general, e.g. fullerenes, graphene, and carbon nanotubes, it is expected that the amount of small molecules coupled to the CNDs changes the optoelectronic properties as well.[35,36] Therefore, it is important to elucidate the type and amount of CND functionalization to be able to design their optoelectronic properties accurately. In 2020, Prato et al. used the Kaiser test (KT) and imine functionalization with a fluorine-containing cinnamaldehyde derivative and subsequent quantitative $^{19}$F-NMR measurements to quantify primary amines in CNDs.[30,37] The same authors compared the amounts of primary amines between the imine functionalization and KT, showing that the primary amine content is higher with the imine method. Bulkier reactants of KT are thought to result in lower accessibility of the primary amines. A common approach to functionalize primary and secondary amines of CNDs comprises the amide coupling due to the high bond strength of amides.[38,39] Since bulky intermediates are present during the amide coupling of small molecules with CNDs, a low amine group accessibility needs to be considered.[40] Other reactions, such as reductive amination, have recently been reported in a quantitative manner to determine the amount of reactive amines.[41] So far, suitable methods to quantify the primary and secondary amines accessible for amide coupling within CNDs are missing.

In this contribution, we propose a method to quantify accessible pendant primary and secondary amine groups in four different CNDs by amide coupling using fluorine-containing model systems. Through different stoichiometries of EDA as starting material, two previously unreported CNDs with varying amounts of primary amines are prepared. Nitrogen-rich CNDs are obtained through a MW-assisted carbonization treatment with two different ratios of Arg and EDA. At the same time, oxygen-rich CNDs were prepared similarly by employing variable ratios of CA and EDA. This selection provides a broad distribution of cases, going from amine rich to amine poor CNDs. Our comprehensive investigation of the elemental and amine quantification provides valuable information about the reactivity and accessibility of the pendant functional groups in the bottom-up CNDs. In addition, from density functional theory (DFT) modeling, we propose that the source of PL in our CNDs is very likely due to small aggregates of molecular fluorophores (MFs).

## Results and Discussion

Four different CNDs were synthesized from small molecules by MW-assisted reactions and purified through dialysis against Milli-Q water. Arg and EDA were used as precursors for one type of CNDs (**Figure 1**), where two different molar ratios of Arg:EDA (1:1.2) (**Arg-1**) and Arg:EDA (1:2) (**Arg-2**) were considered. For the second type of CNDs, we used CA and EDA with molar ratios of CA:EDA (1:1) (**CA-1**) and CA:EDA (2:1) (**CA-2**), respectively. The starting molecules' nature and ratios were chosen to vary the amounts of functional groups, especially the accessible amines, in the prepared CNDs.

After MW-assisted synthesis, the actual presence and properties of the CNDs in the obtained products were investigated by dynamic light scattering (DLS), atomic force microscopy (AFM), UV-Vis absorbance, PL, infrared (IR) spectroscopy and X-ray photoelectron spectroscopy (XPS). The synthesized CNDs' size ranged from 1 to 5 nm, as indicated by DLS (**Figure S1**) and AFM measurements (**Figure 2**, **Figure S2-5**). DLS and AFM characterization showed comparable results, with AFM being likely more reliable due to the difficulty in measuring DLS, as the intrinsic absorption and emission properties of the CNDs were too strong for this technique. Furthermore, for a valid measurement, the actual refractive index of the CNDs must be known, which was only assumed in our case.[42] The size within the DLS is indeed underestimated, and the **CA-2** and **Arg-2** showed a more distinct size distribution, as it could be observed from AFM images. **CA-2** are the smallest CNDs obtained, with an AFM size of about 1.3 nm, whereas the **CA-1** CNDs were the biggest ones, with a size of about 4.5 nm. In addition, larger particles with higher amounts of EDA were observed in both cases. Transmission electron microscopy was not possible to measure with our CNDs, due to the lack of uniformity.[17] These CNDs absorbed light in the ultraviolet region of the electromagnetic spectra and emitted light in the visible blue range (**Figure S6**), with PLQYs between 3% and 12% (**Figure S7**). **Table 1** shows the UV-Vis absorbance maxima, PL maxima, Stokes shift, PLQY, and size values of the four CNDs. A good reproducibility was found by comparing the measured absorbance and PL maxima values of **Arg-1** and **CA-1** with the literature.[15,16] These two species are already reported in the literature, making a comparison feasible.

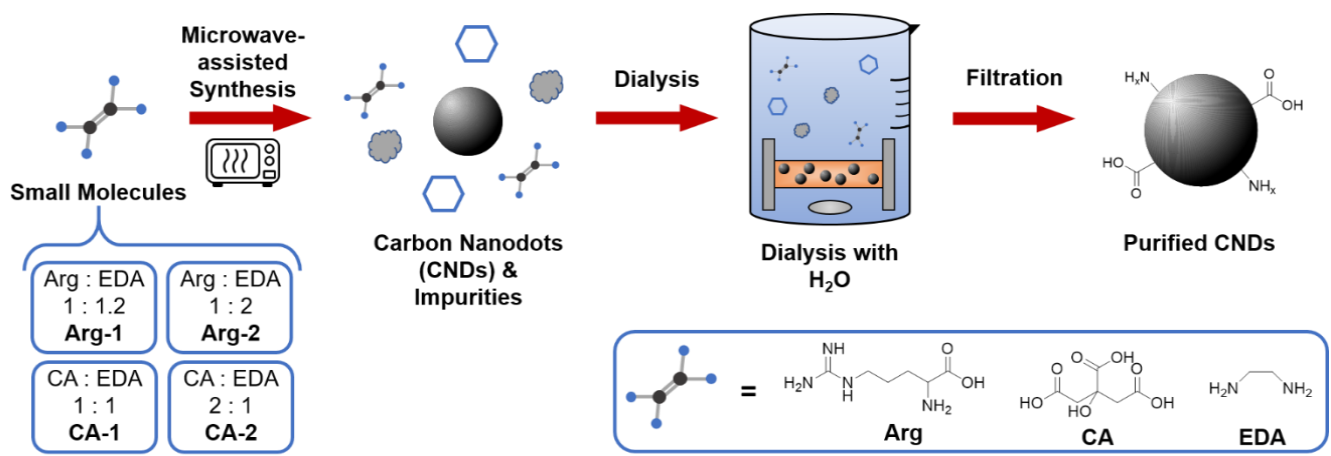

**Figure 1.** The schematic process to obtain and purify CNDs using MW-assisted treatment of Arg, CA, and EDA. For clarity, four different CNDs were synthesized: one type with two ratios of Arg and EDA, and another with two ratios of CA and EDA.

In contrast, the **Arg-2** and **CA-2** species have been synthesized using MW-assisted carbonization and characterized here for the first time.[43,44] These optical data demonstrated no significant change in absorption and emission properties for the two Arg-derived samples. Nevertheless, the two CA-derived CNDs show a considerably different PLQY, with the newly proposed system (**CA-2**) having a four-fold increase in this value, accompanied by a 10 nm hypsochromic shift of the absorption and a higher (~ 20 nm) analogous shift in the PL maximum. Due to their smaller size, a blue shift goes along and reasonably causes the higher PLQY in the **CA-2** sample.

Elemental composition, quantity, and type of functional groups on the CNDs were first analyzed by XPS and Fourier-transform infrared spectroscopy (FT-IR). From the detailed C1s XPS spectra of the CNDs (**Figure S8**), the contribution of oxygenated carbon species (highlighted in red) was lower compared to the starting materials, most likely due to the loss of $CO_x$, $H_2O$, and N-based compounds (i.e., $NH_3$ and $NO_x$) during the MW heating process (carbonization process).[45] In addition, the formation of amide bonds was observed with the addition of EDA. The binding energies of the carboxylic acids shifted to slightly lower values of about 288.7 eV (highlighted in red) compared to the starting molecules. Due to the variety of functional groups, a clear denotation and the exact specification of the proportions was too difficult. At the binding energy of 287.7 eV carbonyl, epoxy, and guanidine-like moieties (as in arginine), can be observed. Between the binding energies of 286.5 eV and 285.7 eV the single-bonded C-N and C-O species, like ether, alcohols, and amines, were found. In the N1s spectra, a shift towards primary amine features with lower binding energies was observed in samples containing higher EDA amounts. In the O1s region of the spectra, a higher amount of EDA resulted in broader signals due to a more diverse chemical environment of the oxygen-containing functional groups.

Primary amines were identified by comparing the IR spectra of the CNDs in the regions around 2900 cm$^{-1}$ with the ones of the precursors (**Figure S9**). Upon addition of EDA, the increase in the intensity of the amide I (1645 cm$^{-1}$) and amide II (1545 cm$^{-1}$) regions indicated a possible reaction pathway between amine from EDA and carboxylic acids from CA or Arg.[46,47] The amide regions are associated with the CO and CN stretching vibration and NH in-plane bending displacement in different proportions. Higher amounts of CA resulted in a more pronounced $v$(C=O) stretching vibration at around 1700 cm$^{-1}$, indicating a higher carboxylic acid content. The $v$(N-H) stretching vibration (2900 cm$^{-1}$) increased with the addition of EDA, whereas at the same time, the $v$(C=O) stretching vibration decreased. This suggests a higher fraction of amine functional groups with increasing EDA amounts and a higher fraction of carboxylic acids with higher amounts of CA, which is in accordance with the XPS detailed spectra.

An augmented amount of CA resulted in a lower amount of nitrogen and a higher amount of oxygen. This was observed in the atomic compositions obtained from elemental analysis (EA) and XPS (**Figure 3**). With increasing EDA amount, the atomic composition was not significantly changed, even though the physical methods like XPS and IR suggested different relative quantities of functional groups. The EA analysis showed that the amounts of hydrogen and oxygen are higher than what could be predicted (**Figure S10**). Since the CNDs are hygroscopic and moisture during the measurement cannot be ruled out, the discussed values are high, most likely due to the presence of residual water, although a careful vacuum drying of the samples was carried out prior to the analysis. In the elemental quantitative information derived from XPS, it was noticeable that the carbon ratio is higher than that derived from classical EA, likely due to adventitious carbon present in the former case (as it is typical during XPS experiments).[48] Since XPS detected no elements other than oxygen, carbon, and nitrogen, we assumed that the percentages of other elements in the sample correspond primarily to oxygen in the EA. Nevertheless, the elemental trends in the EA and XPS chemical characterization as a function of the precursors' ratio variation are comparable and show similar behavior. By comparing the starting materials with the CNDs, it was possible to detect a clear change and reduced heteroatom contents after the MW-assisted reaction.

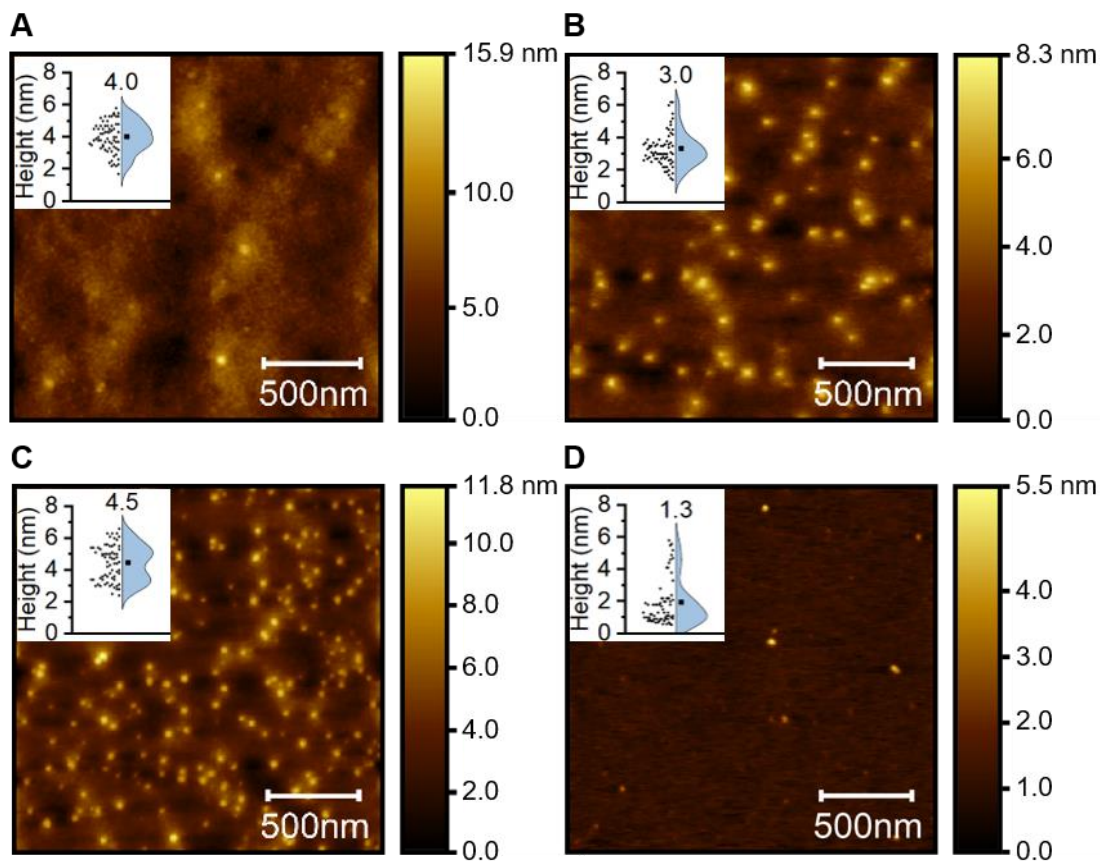

**Figure 2.** 2 μm×2 μm AFM images of (**A**) **Arg-2** (**B**) **Arg-1** (**C**) **CA-1** (**D**) **CA-2** CNDs on a mica substrate. The insets in each figure show the height distribution for each sample, with the median height obtained from ~70 CNDs.

**Table 1.** Measured UV-Vis absorbance ($\lambda_{Abs}$) and PL ($\lambda_{Em}$) maxima compared to literature values (where available), together with the according Stokes shifts, PLQY, and average size obtained from AFM of the four different CNDs.[16,17]

| CNDs | $\lambda_{Abs}$/ nm | Lit.($\lambda_{Abs}$)/ nm | $\lambda_{Em}$/ nm | Lit.($\lambda_{Em}$)/ nm | Stokes shift/ nm | PLQY/ % | Size |
|---|---|---|---|---|---|---|---|
| **Arg-2** | 290 | | 356 | | 66 | 8 | 4.0 |
| **Arg-1** | 290 | 285 | 358 | 356 | 68 | 7 | 3.0 |
| **CA-1** | 348 | 340 | 453 | 450 | 105 | 3 | 4.5 |
| **CA-2** | 338 | | 431 | | 93 | 12 | 1.3 |

Acid-base back titrations were carried out to estimate the total amount of pH-sensitive functional groups present and accessible in the CNDs. First, all functional groups with acidic character were basified with sodium hydroxide, and back titration with hydrochloric acid was performed, thus getting access to all basic and acidic functionalities. By applying the Gran plot,[49,50] it was possible to identify the precise values of the equivalent points, which are directly proportional to the number of functional groups. The number of acid-base sites was quantified by back titration and using the Gran plot analysis (**Figure S11-S14**).[50] The highest acid-base site quantities were observed within the **Arg-2** CNDs, followed by the **CA-2** CNDs (**Figure 4A**). Considering the isoelectric point (pI) of **CA-2** at pH 4.0, the sample must have a high amount of carboxylic acids or other acidic moieties (**Figure S15**). The higher pI at pH 9.4 for the **Arg-1** CNDs indicated predominantly basic moieties. **Arg-2** instead showed the pI at pH 7.6, suggesting nitrogen heterocycles with typically low pKa values. For comparison, the pI of arginine is at pH 10.76 and for proline at pH 6.3.[51] This correlates well with the lower proportion of basic amine functional groups in this species found by the other chemical analysis methods.

Thermogravimetric analysis (TGA) was performed for all samples in air and nitrogen up to 900 °C (**Figure S16**). The thermograms can be divided into three general regions: (I) From 100 °C to 300 °C, an initial thermal decomposition of thermolabile functional groups was observed, (II) the mass loss from 300 °C to 600 °C (in $N_2$) indicated the decomposition of stable functional groups present in the bulk or multiple C-hetero atom bonds, (III) the decay from 450 °C to 650 °C (in air) is related to the combustion of condensed carbonaceous structures. **CA-2** showed a higher mass loss in the first region than the **CA-1** CNDs, indicating a larger amount of functional groups. For Arg CNDs, the loss of thermolabile groups started earlier than for CA CNDs, and in general, the loss in this region was more significant. **Arg-2** showed the highest loss among all four samples. Thermolabile functional groups decomposing between 100 °C and 300 °C followed the same tendency of the acid-base functional group content obtained through back titration. Regarding the morphology, one can observe higher degrees of carbonized residues for the CA CNDs in the TGA analysis measured in an inert atmosphere. This reflects the visual appearance of the dark brown CA CNDs compared to the orange-to-brownish Arg CNDs (**Figure S17**). In addition, the Arg CNDs showed lower decomposition temperatures. Thus, they are less thermally stable than the CA CNDs.

To probe the presence and amount of accessible primary amines, KT was employed. When the results of KT were compared with the EA, it appeared evident that, even with a similar amount of nitrogen, the **Arg-2** CNDs show the highest accessible primary amine content of 1900 µmol·g$^{-1}$ (**Figure 4**). Consistent with this behavior, it has been reported in the literature, using carbon isotope-labeled starting materials and subsequent $^{13}$C-NMR studies, that EDA is the major contributor to the shell of the CNDs, while Arg is mainly responsible for the core.[15] Therefore, the amount of accessible functional groups will change with the amount of EDA added, even though the overall amount of nitrogen remains similar. This resulted in 33% (in molar fraction) more primary amines in **Arg-2** CNDs than in the **Arg-1** CNDs. The amount of EDA precursor employed is 39% more in **Arg-2** than in **Arg-1,** consistent with the detected higher primary amine content of the former compared to the latter. The obtained accessible primary amine amount of the **Arg-1** CNDs was almost identical to the reported literature value of 1350 µmol·g$^{-1}$.[30] The EDA precursor amount used for the synthesis of the **CA-1** CNDs in comparison to the **Arg-1** CNDs is 23% lower but lead to an only 7% lower content of primary amines (1200 µmol·g$^{-1}$). This suggests that either more primary amine groups can be found on the **CA-1** CNDs, or they are more accessible for the KT reagents than in the **Arg-1** CNDs. However, when the amount of EDA was halved, the amine content dropped to 50 µmol·g$^{-1}$ for the **CA-2** CNDs, which was 96% lower than the **CA-1** CNDs. As a result, EDA induced different product formations for the resulting CNDs when changing the ratio of the starting materials. On the other hand, when a higher amount of EDA was used to prepare the Arg-based CNDs, the quantity of primary amines increased rather linearly compared to the CA-based CNDs.

Since it was shown that KT cannot access the total amount of primary amines due to a possible hindrance effect, the CNDs were chemically functionalized with 4-fluorocinnamaldehyde to form an imine.[30] Followed by quantitative $^{19}$F-NMR measurements with an internal standard, this method enabled fluorine functionalized primary amine groups to be quantified. A value of 3960 µmol·g$^{-1}$, for **Arg-1**, showed good agreement with the literature value of 4100 µmol·g$^{-1}$. In our work, we also found higher values for the imine functionalization compared to KT and a similar trend from high to low EDA contents (**Figure 4A**). In the case of the CA CNDs, the quantity of primary amines obtained from the KT and the imine functionalization showed a lower deviation than the other CNDs, suggesting a higher accessibility of the moieties. The **CA-2** CNDs showed a ten times higher primary amine content from the imine functionalization than the KT result. This indicates that different products likely form during the synthesis of the CA CNDs, although the employed precursors are the same (only the relative ratio varied). In agreement with the literature, EDA changed the surface composition of the CNDs and, therefore, the content of accessible functional groups as well.[15]

Amide bonds were chosen for the future functionalization of the CNDs due to their higher bond strength compared to imines, and the utilization of amide coupling was initially studied by quantitative $^{19}$F NMR. The primary and secondary amine groups were functionalized through a peptide coupling reaction with 4-fluorobenzoic acid in the presence of EDC and NHS (see details in the experimental section) (**Figure 4C**). To prevent self-cross-linking of the CNDs, they were added to the reaction mixture after activation of the 4-fluorobenzoic acid with EDC. After extracting the coupling product, the $^{19}$F-NMR was measured in DMSO-d6 with 4,4'-difluorobenzophenone as an internal standard. Contrary to expectation, the primary and secondary amine content followed a similar trend to the back titration. The highest degree of functionalization was obtained with the **Arg-2** CNDs, probably due to a higher reactivity and accessibility of amine groups in the Arg CNDs. Considering the back-titration access to the total acid-base moieties, the CNDs can be functionalized in the 1% to 6% range by amide coupling, under the conditions studied herein. Assuming that the functionalization of the primary amines is fully effective during the imine reaction, the maximum degree of functionalization of the acid-base sites is 39% for the **Arg-1** and 21% for the **CA-1** CNDs. In this comparison, it is necessary to consider the different starting materials of the coupling reactions since 4-fluorocinnamaldehyde has an alkene chain with three carbons between the reacting aldehyde and the ring. On the other hand, the reacting carboxylic acid is located directly on the benzene ring of the 4-fluorobenzoic acid, and the acid is activated by EDC, which is a bulky molecule.[40] It seems to make no difference that the secondary amines can be additionally functionalized with the amide coupling. Most likely, this is due to the lower activity and accessibility of secondary amines compared to primary amines. Nevertheless, the amide coupling with 4-fluorobenzoic acid better reflects the functionalization that could be carried out with other bulky molecules, for example, to produce dye-CND hybrids useful for optoelectronic applications, and one can expect similar degrees of functionalization in such cases.

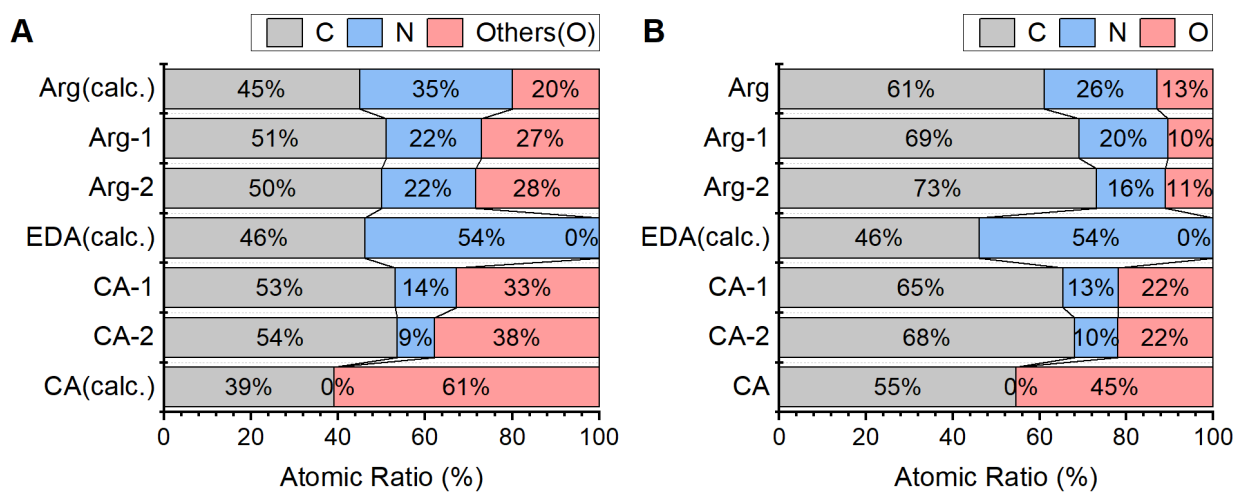

**Figure 3.** (**A**) EA data for the four synthesized CNDs compared with the theoretical atomic ratios of the starting materials. (**B**) Atomic composition of the CNDs, CA, and Arg derived by XPS spectra compared with the theoretical atomic ratios of EDA.

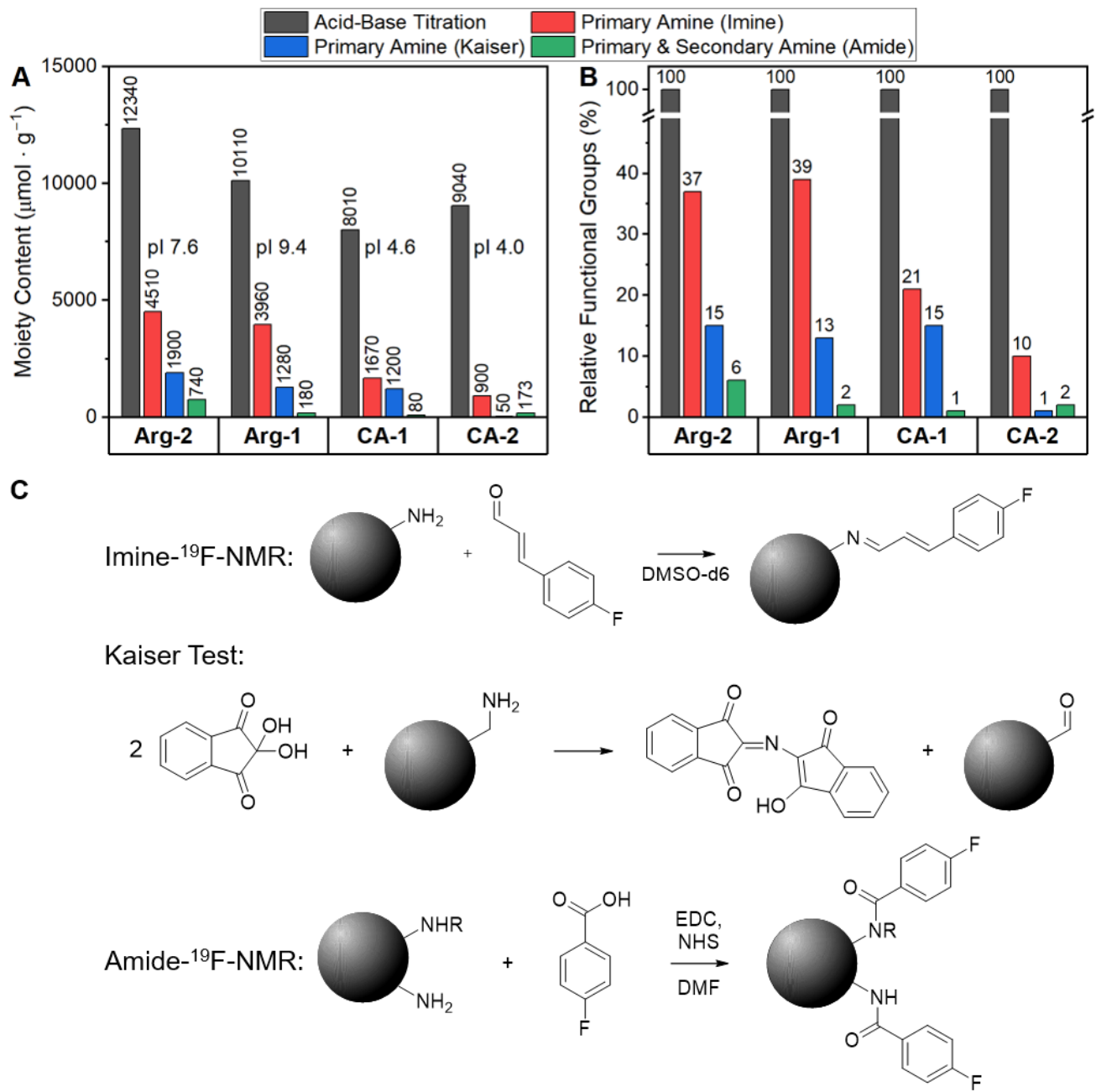

**Figure 4.** (**A**) Accessible functional groups determined through different methodologies. Total acid-base groups by back-titration, primary amine by KT, primary amine by quantitative ¹⁹F-NMR of imine functionalized, and primary and secondary amine by quantitative ¹⁹F NMR after the amide functionalization. (**B**) Functional group content relative to the total amount of acid-base groups. (**C**) The reaction schemes of KT, imine, and amide couplings are depicted below for the sake of clarity.

To understand the relationship between the unique structure and optical properties of our CNDs, (time-dependent) density functional theory ((TD)-DFT) calculations were performed to study the electronic structure and to compute absorption/emission spectra of CNDs. Unfortunately, the structure-photoluminescence relationship of CNDs is yet to be established, thus, it is challenging to unambiguously report the source of PL in any CDs synthesized by hydrothermal treatment. Therefore, our modelling was based on *ad-hoc* structures which could be possibly formed based on the synthetic protocols/reports in the literature. These model structures have compositions as close as possible to the experimental XPS data. However, the models do not perfectly correspond to the experimental composition, due to the inherent complexity of the studied systems. Thus, with the selected models we aim to separate the complex CNDs composition into different parts, considering either the size of the system, its nitrogen/oxygen content and the possibility of molecular fluorophore assembly as reliable building blocks. Starting with **Arg-1** CNDs, three families of model systems (**Figure 5**) were designed based on: I) available experimental information ('large' model), II) data in the literature ('medium' model),[52] and III) possible formation of small organic molecular fluorophores (MFs) during the synthesis of CDs ('MF' model).[53,54]

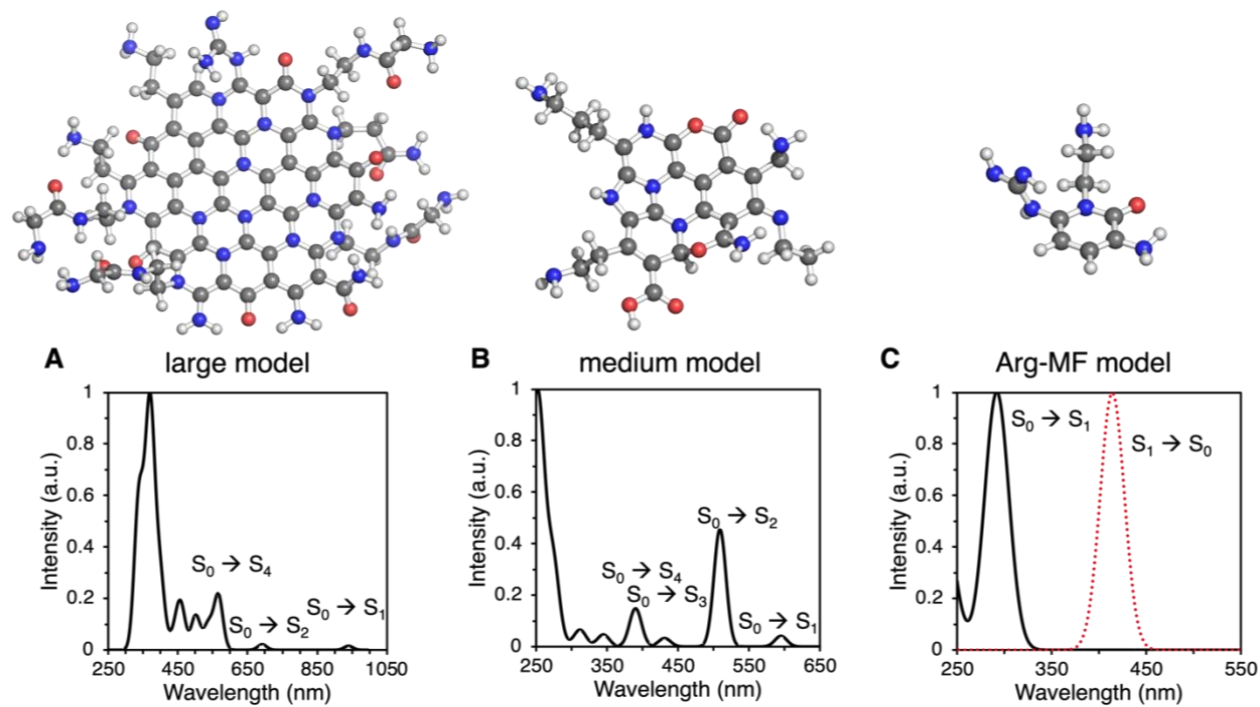

**Figure 5.** Top panel: structures of the different families of model systems considered for the (TD-)DFT calculations. Coloring: carbons-grey; nitrogens-blue; oxygens-red; hydrogens-white. Bottom panel: Calculated absorption (black) and emission (red) spectra of (**A**) large, (**B**) medium, (**C**) Arg-MF models. Line spectra were convoluted by a Gaussian function assuming the inhomogeneous broadening of peaks with $\sigma$ = 30 nm.

Our calculations demonstrated that among the three selected models, only those based on MFs were in good agreement with the experimental data, while the other two led to a strong red shift in the absorption spectra (**Figure 5**, **Table S1**), implying that the PL would be in the IR region. In particular, the calculated vertical excitations of the 'large' model demonstrated that the first excited state $S_1$, lying at 941 nm, is strongly red shifted compared to experiments ($\Delta\lambda$ = 651 nm, $f$ = 0.021), implying that the experimentally measured absorption at 290 nm does not originate from large N, O-functionalized N-doped polycyclic aromatic hydrocarbons. It shall be noted that beside surface functional groups, graphitic nitrogens were also considered to achieve higher N-doping in our model. Although it has been reported that graphitic nitrogen red-shifts the absorption spectra, we observed that the 'large' model with no N-doping of the aromatic core also exhibits the red-shift in the absorption spectrum in comparison with the experimental data (Figure S18). Similarly, a large red-shift of 305 nm was observed for the 'medium' model, with $S_0 \rightarrow S_1$ transition peaking at 595 nm, even though this exact model structure was suggested in the literature as the source of PL of Arg-CNDs.[52] On the other hand, when the arbitrary model of Arg-MF, constructed based on the possibility of a synthetic reaction to occur under high temperature conditions in the complex, is considered as a constituent for CNDs, the $S_0 \rightarrow S_1$ excitonic transition (**Figure S19**) has been found to have the first bright absorption peak at 292 nm (**Figure 5, Table S1**), which differs only by 2 nm from the experimental data. In addition, its emission is red shifted by 56 nm compared to experimental data (414 nm vs. 358 nm measured). Comparing these three families of models, it could be rationalized that MFs can be used to also obtain CNDs synthesized in the condition of high temperature, but not high enough to ensure complete carbonization, which may not be the general rule in CNDs, but we believe it occurs in our synthesized CNDs.[3,55]

Nevertheless, this Arg-MF model is too small to account for the size distribution of CNDs reported in **Figure 2** (**Figure S1–S5**). Therefore, additional computations were performed considering an extended model in which one (Arg-1-MF-a) or two (Arg-1-MF-b) more molecules of Arg and EDA were covalently linked together with a short aliphatic chain (**Figure 6**). Interestingly, increasing the size of the Arg-MF model has a negligible effect on their absorption and emission spectra (**Figure 6A**, **B**, **Table S2**), as the bright absorption peak corresponding to the first (second) excited state lies at 308 nm (293 nm) for Arg-1-MF-a (Arg-1-MF-b). This is due to the fact that these transitions can be described as π-π*, both localized on the aromatic cores (**Figure S20**), which are in turn separated by a short aliphatic chain. To further strengthen our idea of CNDs made from MFs, MD simulations with five non-covalently bound Arg-1-MF-b molecules were carried out.

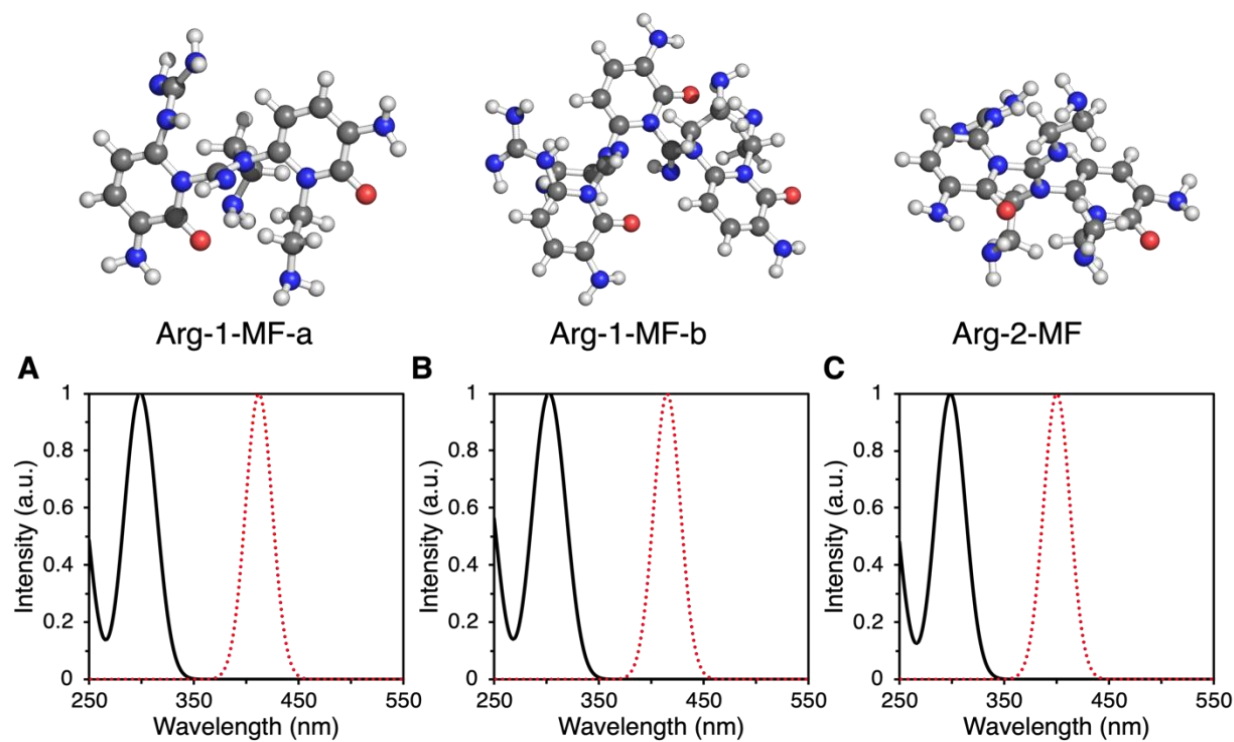

**Figure 6.** Top panel: Three different structure models based on small organic molecules, which explain the optical properties of Arg:EDA CNDs. The Arg-MF model can be further extended to dimers (Arg-1-MF-a) and trimers (Arg-1-MF-b), with similar absorption/emission agreeing with the measured experimental spectra. Coloring: carbons-grey; nitrogens-blue; oxygens-red; hydrogens-white. Bottom panel: Calculated absorption (black) and emission (red) spectra of (**A**) Arg-1-MF-a, (**B**) Arg-1-MF-b, (**C**) Arg-2-MF models. Line spectra were convoluted by a Gaussian function assuming the inhomogeneous broadening of peaks with $\sigma = 30$ nm.

Our results demonstrated that these molecules formed a stable assembly (**Figure S21**), where the average Coulomb and Lennard-Jones interaction between these fluorophores is –41.8 kcal·mol$^{-1}$ and –42.7 kcal·mol$^{-1}$, respectively. The calculated radius of gyration averaged over the entire simulation trajectory has a value of 1.1 nm, which agrees well with the measured hydrodynamic diameter (**Figure S1**). The bimodal broad distribution of this plot suggests that our assemblies may be highly dynamic.

Additionally, the calculated excitation and emission values for Arg-2-MF model showed no significant shifts compared to the above considered Arg-1-MF models (**Figure 6C**). This arbitrary model is used to rationalize both absorption and emission spectra of the **Arg-2** CNDs, as it contains an additional EDA molecule covalently bound to the part connecting two aromatic arginine cycles, thus representing the higher 1:2 ratio of Arg:EDA present in **Arg-2**. Adding the EDA molecule in the Arg-2-MF model increases the number of amine groups in the Arg-CNDs, which agrees with the KT experiments (**Figure 4A**).

Considering that out of three different families the small MF models performed the best in reproducing the experimental optical data of the Arg:EDA CNDs, the MF based models were also applied to the CA:EDA CNDs. Moreover, it has been accepted in the literature that a MF called 5-oxo-1,2,3,5-tetrahydroimidazo-[1,2-α]-pyridine-7-carboxylic acid (IPCA) (**Figure 7**) is be formed during the high-temperature synthesis used for CNDs of about 200 °C.[56–58] Our theoretical calculations of IPCA agreed well with literature,[59] showing that the first absorption peak lies at 346 nm and emission at 432 nm, i.e., 2 nm blue-shift and 21 nm blue-shift, respectively, in comparison with the experimental value of **CA-1** CNDs (**Figure 7A**, **Table S3**). Thus, for the synthesized **CA-1** CNDs, the origin behind the PL could be explained by the presence of aggregates of MF IPCA in the sample.[60,61] On the other hand, computational models based on covalently bonded IPCAs (CA-2-MF-a) and a model composed of short oligomers (CA-2-MF-b) were considered for the **CA-2** CND, as what we have there is more CA than EDA groups. Both these **CA-2** models demonstrated excellent agreement with experiments, where the absorption (emission) peak of the CA-2-MF-a model at 336 nm (430 nm) exhibited a negligible 2 nm (1 nm) blue shift compared to the experimental data (**Figure 7B**, **Table S3**, **Figure S22**). Moreover, our computations with the CA-2-MF-b model showed that **CA-2** CNDs could be composed of small oligomers with covalently attached IPCAs (**Figure S23**), being the source of PL, as the first absorption peak corresponding to $S_0 \rightarrow S_1$ and $S_0 \rightarrow S_2$ vertical transitions is blue shifted from the experimental data by 19 nm. These transitions are excitations localized on the IPCA part of the oligomer (**Figure 7C**, **Table S3**, **Figure S24**).

The explanation of optical properties based on MFs is supported by the fact that during the dialysis purification of the CNDs, high fluorescence of the dialysate under UV irradiation was observed (**Figure S25**); additionally, in the publication by Baker et al.[62] it was shown that the quantum yield decreases with further purification of the CNDs, which is attributed to highly fluorescent small molecules entering the dialysate during purification. Nevertheless, we would like to once again stress that the source of PL in CNDs is a difficult task to tackle, and the theoretical models shall be considered as a simplification of these complex systems.

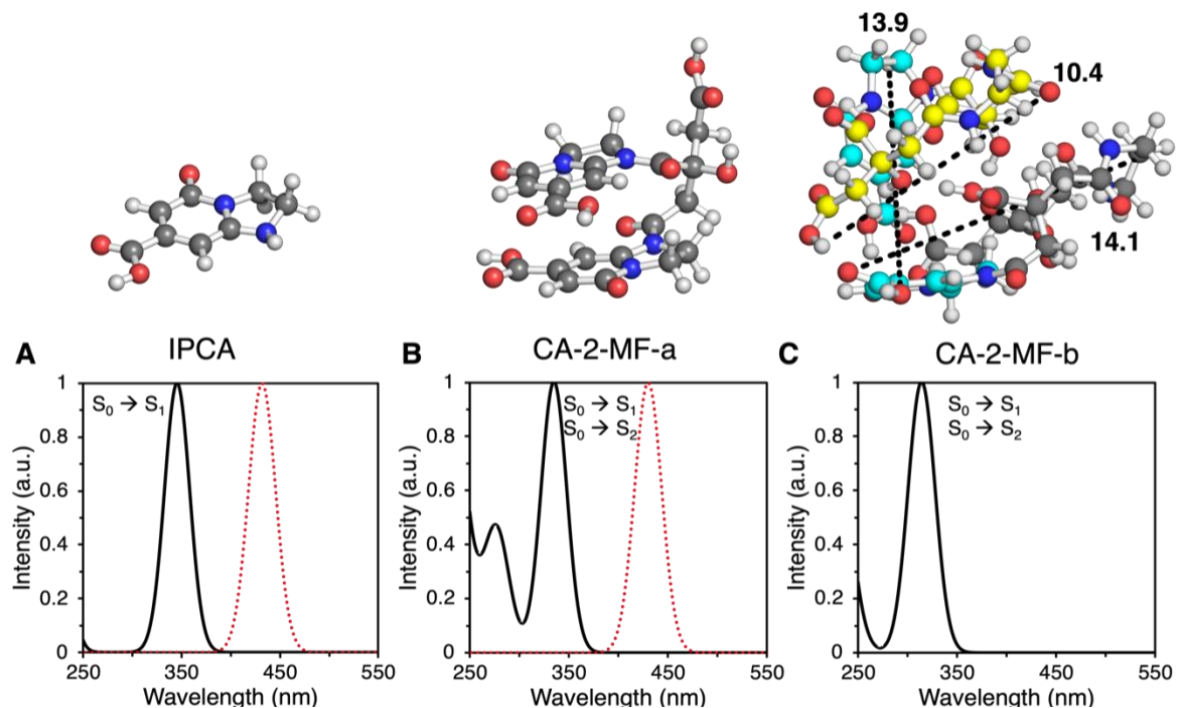

**Figure 7.** Top panel: Structure models based on small organic molecules, which explain the optical properties of CA:EDA CNDs. Coloring: carbons-grey, cyan, yellow; nitrogens-blue; oxygens-red; hydrogens-white. For CA-2-MF-b model, carbons of two non-bonded short oligomers are colored grey and yellow. The carbons in cyan represent IPCA covalently bonded to the oligomer. Distances are displayed in Å. Bottom panel: Calculated absorption (black) and emission (red) spectra of (**A**) IPCA, (**B**) CA-2-MF-a, (**C**) CA-2-MF-b models. Line spectra were convoluted by a Gaussian function assuming the inhomogeneous broadening of peaks with $\sigma$ = 30 nm. Due to the size of the system, def2 SVP basis set was used for CA-2-MF-b model, which could lead to a blue shift of around 10 nm (see the footnote in **Table S3**).

## Conclusion

We report the synthesis, detailed characterization, and quantification of accessible functional groups of two novel and two previously reported bottom-up CNDs obtained by MW-assisted synthesis. Through the variation of the starting material and the addition of different amounts of EDA, we were able to tune the quantity and type of functional groups. IR, XPS, and EA indicate changes of the functional group amount with varying EDA quantities. The functional groups were in depth quantified by KT, acid-base back titrations, and functionalization with fluorine containing model systems with subsequent quantitative $^{19}$F-NMR measurements.

We used a modified method to determine the degree of functionalization after the amide coupling of small molecules with CNDs. We show that a functionalization from 1% to 6% can be achieved with the amide coupling, even though the amine content is probed to be higher. Moreover, the CNDs do not follow the expected trend from the KT or imine functionalization for the functionalization by amide coupling. This demonstrates that the amount of amine groups cannot be used as a direct measure for determining the potential for covalent functionalization of CNDs. Determining the degree of accessibility serves as a tool for targeted functionalization to produce precisely tailored optoelectronic materials. CNDs' accessibility and reactivity, the coupling reaction, and the molecule used to functionalize the CNDs, strongly influence the degree of functionalization.

In addition, we used different model systems and computed their optical properties. The calculated spectra based on simplified models of MFs and the related oligomers show great agreement with the experimental data, while for larger aromatic systems the calculations result in a strong red shifted absorption and emission. This indirectly validates the assumption that the formation of small molecules during the synthesis of the CNDs is the main cause of the photophysical properties in these still elusive nanocarbons.

## Materials and Methods

D/L-Arginine (Arg, ≥ 98.5%), citric acid (CA, ≥ 99.5%), 3-(((ethylimino)methylidene)amino)-N, N-dimethylpropan-1-amine hydrochloride (EDC, ≥ 99%), and ethylenediamine (EDA, ≥ 99.5%) were purchased from Carl-Roth. 4,4'-difluorobenzophenone (99%), N-hydroxysuccinimide (NHS, ≥ 99%) were purchased from Sigma-Aldrich. 4-fluorobenzonitrile (99.0%) were purchased from Fluorochem. Deuterated dimethyl sulfoxide (DMSO-d6, 99.8%) were purchased from Deutero. Anhydrous N, N-dimethylformamide (DMF, 99.8%, AcroSeal) was purchased from Thermo Scientific. Dimethylsulfoxide (DMSO, 99.5%) was purchased from Gruessing. Ultrapure water was obtained with the Milli-Q Direct Water Purification System. Unless otherwise stated, the reagents were used without further purification. The MW-assisted reactions were performed in a Silvia homeline and CEM Discover-SP MW set-ups. The purification of CNDs was performed with Spectra/Por Biotech-Grade Cellulose Ester (CE) dialysis tubing with a molecular weight cut-off (MWCO) of

500 Da–1000 Da. NMR-spectra were recorded with the Bruker Avance II 400 MHz, Bruker Avance III 400 MHz HD, and the Bruker Avance III HD 600 MHz. XPS measurements were conducted with a PHI 5000 VersaProbe IV Scanning ESCA Microprobe (Physical Electronics) with monochromatized Al K$_α$ 1486.6 eV X-ray source (beam diameter 100 µm, X-ray power: 25 W, X-ray Voltage: 15 kV). Samples were prepared by filling a sample cap made of Teflon and attaching them to the XPS sample holder using an isolating tape. Time steps of 100 ms, a step size of 0.2 eV, and an analyzer pass energy of 27 eV were used for measuring the detail spectra. For every detailed region 30 sweeps or a P/N ratio equal to 180, were set. For the general survey a time per step of 50 ms, analyzer pass energy of 224 eV, and a step size of 0.8 eV, were used. The sample surface was charge neutralized with slow electrons and argon ions, and the pressure was in the range from $10^{-7}$ Pa to $10^{-6}$ Pa during the measurement. Data analysis was performed using the CasaXPS software.[63] All of the spectra were calibrated using C1s signal located at 284.4 eV, associated to C-C bond and the signals were decomposed using GL(30) function. DLS and zeta potential measurements were conducted with a Malvern Zetasizer Nano-ZS device (Malvern, U.K.) at 20 °C. The system was operated by the Zetasizer software from Malvern Panlytical (Malvern, U.K.). UV-Vis measurements were conducted with the Bio-Tek Uvikon XS Spectrophotometer and the system was operated with the software UV VISion Lite. Fluorescence measurements were conducted with a Jasco FP-8300 Fluorescence Spectrometer. EA (CHN) was performed with the Thermo FlashEA – 1112 Series equipment. Back titrations were conducted with the TitroLine 7000 from SI Analytics. ATR-IR was performed on a VERTEX70/ Platinum ATR (Fa. Bruker Optics) with a minimal resolution of 0.2 cm$^{-1}$. AFM was performed with the AIST-NT SmartSPM 1000 under ambient conditions in AC mode. AFM probe (SHR150 probe, BudgetSensors', single hydrophobic diamond-like carbon extratip at the apex of Gold coated silicon etched probe, 150 kHz, 5 N m$^{-1}$) Data analysis was performed with Gwyddion software.[64] Thermogravimetric measurements were carried out on a TGA Q5000IR (TA instruments, New Castle, DE, USA). The analyses were performed using a 100 µL Platinum-HT TAG pan. In order to remove adsorbed moisture, each sample was subjected to a sequence of isotherms under N$_2$ (at 60 °C for 10 minutes, then at 80 °C for 3 minutes and then at 100 °C for 2 minutes), immediately before the measurement ramp (10 °C·min$^{-1}$ from 100 °C to 900 °C) either under N$_2$ atmosphere or under air atmosphere.

**CNDs Synthesis**

**Arg + EDA**

For the sake of clarity, the details of the already reported and slightly modified method reported by Prato et al.[52] are given here. In a MW vessel, Arg (87.4 mg, 0.502 mmol, 1 eq) was dispersed in 100 µL Milli-Q water and EDA (**Arg-1**: 33 µL, 0.61 mmol, 1.2 eq); (**Arg-2**: 67 µL, 1.0 mmol, 2 eq) was added. With an added stirring bar, the reaction mixture was stirred for 30 s at 300 rpm on a magnetic stir plate. The vessel was placed in the MW reactor (CEM Discover SP) with 12 power cycles including a power interval of 200 W for 15 s and a cooling interval of 5 s. Overall reaction time was 180 s. The temperature was limited between 240 °C and 250 °C. The obtained brown resin was dispersed in 3 mL Milli-Q water and filtered over a 0.2 µm PTFE syringe filter.

**CA + EDA**

For the sake of clarity, the details of the already reported and slightly modified method reported by Haynes et al.[16] are given here. CA (1.4410 g, 7.5004 mmol, 1 eq) was dissolved in 5 mL Milli-Q water and was stirred until everything was dissolved. EDA (**CA-1**: 0.50 mL, 7.5 mmol, 1 eq); (**CA-2**: 0.25 mL, 3.8 mmol, 0.5 eq) was added and the mixture was stirred for 20 minutes. The reaction mixture was heated in a domestic MW oven for five minutes at 400 W (50% power). The obtained brown resin was dispersed in 10 mL Milli-Q water and the brown dispersion was filtered over a 0.2 µm PTFE syringe filter.

The filtered dispersions of all the CNDs were filled up to a volume of 10 mL and dialyzed in a 0.5 kDa – 1 kDa dialysis membrane against Milli-Q water for 48 hours with water changed every two hours during the day. The obtained brown dispersion was lyophilized and an orange/ brown solid was obtained (**Figure S17**).

$^1$H-NMR for each CND species showed broad signals to ensure the presence of CNDs (**Figure S26**).[65]

**Amide Coupling of CNDs with 4-fluorobenzoic acid**

4-Fluorobenzoic acid (41.3 mg, 0.295 mmol, 1 eq) was dissolved in 20 mL anhydrous DMF. EDC (111.3 mg, 0.5806 mmol, 2 eq) and NHS (65.4 mg, 0.568 mmol, 2 eq) were added and the mixture was stirred for 15 minutes at room temperature. CNDs (25.2 mg) were finally added and the reaction mixture was stirred overnight at room temperature. The solvent was removed under reduced pressure and the obtained crude was dispersed in 10 mL aqueous hydrochloric acid (0.1 M). The mixture was washed three times with chloroform and the obtained aqueous phase was dialyzed in a 0.5 kDa–1 kDa MWCO dialysis tube against Milli-Q water for 24 hours with water changed every two hours during the day. The obtained brown dispersion was lyophilized and an orange/brown solid was obtained.

**Imine Formation of CNDs with *trans*-4-fluorocinnamaldehyde**

CNDs (10.3 mg) were dispersed in 0.6 mL deuterated DMSO and trans-4-fluorocinnamaldehyde (18.3 mL, 0.140 mmol) was added. The reaction mixture was stirred overnight at room temperature and directly analyzed.[30]

**Kaiser Test**

Four different solutions were prepared for the modified Kaiser test.[66]
1. Ninhydrine solution: 2.5 g of ninhydrine is dissolved in 50 mL of absolute ethanol.
2. KCN solution: 2 mL of 0.03 M aqueous KCN solution is diluted to 100 mL with freshly distilled pyridine.
3. Phenol solution: 40 g of phenol is dissolved in 10 mL ethanol by gently heating the solution until it is dissolved.
4. Ethanol solution: 60 mL ethanol are diluted with 40 mL of Milli-Q water.

In a 4 mL screw cap vial around 1 mg of CNDs were placed together with 75 µL of the phenol solution, 100 µL of the KCN solution, and the mixture was sonicated for two minutes. Then 75 µL of the ninhydrine solution was added and the mixture was stirred at 120 °C for 10 minutes. After cooling to room temperature, the reaction mixture was diluted with 32 mL of the ethanol solution and a UV-Vis absorbance spectrum was measured. As a reference for the absorbance measurement, the same procedure was performed without the addition of CNDs to the solutions. Each CNDs sample underwent this preparation at least three times and the average value was calculated.

**Acid/base Back Titration**

An HCl (0.10987 M) and NaOH (0.6024 M) solutions were prepared, and the titer was determined with a 0.1 M NaHCO$_3$ solution. For quantification, around 9 mg of the CNDs was weighed, dissolved in 7 mL Milli-Q water, and 1 mL of the NaOH solution was added to the mixture. Back titration was performed with the prepared HCl solution (0.10987 M). The pH was measured throughout the titration. A Gran plot was performed with the obtained values of the added volume of HCl solution and the pH to determine the two equivalence points through linear fitting.[50] Here, the equivalence points are within a small ±15% volume interval around pH 7. By subtracting the two equivalence points, the number of active sites within the CNDs are obtained. The first equivalence point results from the neutralization of the excess hydroxide ions and the second from the complete protonation of the functional groups of the CNDs.

**Computational Methods**

To examine the possible models explaining the optical properties of the experimentally synthesized Arg:EDA and CA:EDA CNDs, the lowest excited states were analyzed in terms of the vertical (de-)excitation energies and the corresponding oscillator strengths as well as in terms of natural transition orbitals (NTOs). All calculations were performed for hydrated systems by employing the implicit universal solvation model based on solute electron density (SMD)[67] and using the Gaussian16 package.[68] The geometries of all the models were fully optimized without any symmetry constraints, within the wB97X-D/def2-SVP level of theory.[60,69] Genuine minima at the potential energy surface were confirmed by the absence of imaginary frequencies in the harmonic vibrational analysis. The electronic excited levels were computed at the time-dependent density functional theory (TD-DFT)[70] framework with the range-separated hybrid wB97X-D exchange-correlation functional and def2 TZVP[60] basis set on all atoms. The electronic vertical excitation energies (VEEs) were calculated using the linear response (LR) with the first twenty singlet states taken into consideration.[71–73] The vertical emission energies were obtained by first optimizing the S$_1$ geometry with the def2-SVP basis set, followed by the calculation of emission energies with the def2-TZVP basis set applying the equilibrium solvation regime for the excited state calculations and nonequilibrium solvation for the subsequent calculation of the ground state. NTOs were rendered with the Chemcraft software (version 1.8).[74]

In molecular dynamics (MD) simulations, all models were parametrized in the Generalized Amber Force Field (GAFF)[75] RESP[76] partial charges were assigned by the Antechamber tool[77] from the Amber software package[78] after the geometry optimization of single molecule (Arg-1-MF-b) at the B3LYP/6-31G* level,[79,80] and calculation of the electrostatic potential at this geometry at the HF/6-31G* level in gas. Five molecules were randomly placed in a cubic simulation box of 3.5 × 3.5 × 3.5 nm$^3$ and solvated with TIP3P[81] explicit water molecules (1242 in total). After minimization, a two-step equilibration was carried out. First, the system was thermalized from 0 K to 300 K for 1 ns with the V-rescale[82] thermostat with a 0.1 ps scaling constant. Then, 2 ns equilibration of pressure was performed using the isotropic Berendsen barostat[83] to keep the pressure at 1 bar during the simulation, the time constant for pressure relaxation was set to 2 ps, and the temperature was kept at 300 K. 100 ns long production MD simulations were carried out in the canonical (NPT) ensemble with a 2 fs time step. The bonds involving hydrogen were constrained using the LINCS[84] algorithm. The temperature was kept at 300 K with the V-rescale thermostat with a 0.1 ps scaling constant, the pressure was kept at 1 bar using the isotropic Parrinello–Rahman barostat[85] and the time constant for pressure relaxation was 1 ps. The electrostatic interactions were treated by means of the particle-mesh Ewald (PME)[86] method with a real-space cut-off of 1 nm; the same cut-off was applied for van der Waals interactions. Periodic boundary conditions were applied in all three dimensions. All MD simulations were performed in Gromacs 5.0.[87] Figures from the MD simulations were rendered in PyMoL.[88]

To assess the pH effect over the absorption properties on the investigated systems, different protonation states for the guanidine/amine groups have been considered for the Arg:EDA CNDs, as well as for the carboxylic groups for the CA:EDA CNDs. The absorption spectra of the models with protonated guanidium groups showed the same absorption position as in the spectra of neutral structures (Figure S22), due to the localization of the excitonic transitions on the aromatic parts of the molecular fluorophores for both holes and electrons. On the other hand, for the CA:EDA systems, it is known from literature that the deprotonation of the carboxylic group of IPCA causes the blue-shift of the absorption spectra. The calculations with different number of deprotonated carboxylic groups in IPCA and CA-2-MF-a models display a blue-shift in the absorption spectra in comparison with the experimental data only when all carboxylic groups are deprotonated (Fig. S25). At the experimental pH condition, we can expect either one or two deprotonated carboxylic groups. For this situation, a negligible shift in the absorption spectra has been observed. We can thus conclude that i) the CA:EDA molecules have either protonated groups due the embedding of IPCA molecules within the CNDs framework while not being exposed to the solvent, or ii) different protonation states co-exist, leading to a broader absorption peak centered around 330 nm (green curve in the Figure 25e).

## Acknowledgements

We thank the financial support of the DFG and of the National Science Center, Poland through the joint project "Low-Dimensional Nano-Architectures for Light Emission and Light-to-Electricity Conversion" (LOW-LIGHT, grant no. UMO/2020/39/I/ST4/01446). S.O. also acknowledges the financial support of the "Excellence Initiative – Research University" (IDUB) Program, Action I.3.3 – "Establishment of the Institute for Advanced Studies (IAS)" for funding (grant no. UW/IDUB/2020/25). T.G. also acknowledges the financial support of the European Research Council through the ERC StG project JANUS BI (grant agreement No. [101041229]) and of the European Commission through the H2020 FET-PROACTIVE-EIC-07-2020 project LIGHT-CAP (grant agreement No. [101017821]). This research was carried out with the support of the Interdisciplinary Center for Mathematical and Computational Modeling at the University of Warsaw (ICM UW) under grants no. G83-28 and GB80-24. E.M. would like to acknowledge Centro Studi di Economia e Tecnica dell'Energia Giorgio Levi Cases (project PRINTERS) and University of Padova (P-DiSC#05BIRD2021-UNIPD) for funding. The authors thank Heike Hausmann (Giessen) for performing the NMR experiments, Pascal Schweitzer (Giessen) for the support with the AFM measurements, and Samuel Pressi (Padua) for technical assistance with the TGA experiments.

Supporting Information
The authors have cited additional references within the Supporting Information.[30,50,51,64,66,89–94]

**Entry for the Table of Contents**

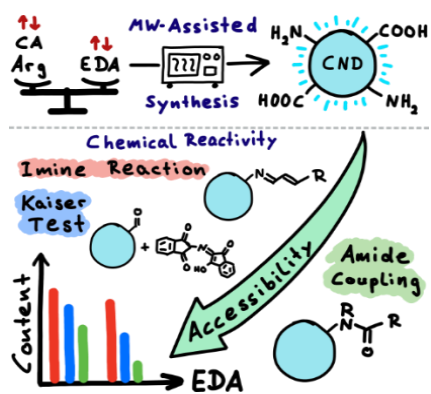

We report a microwave(MW)-assisted synthesis of four different carbon nanodots (CNDs), starting from arginine (Arg) or citric acid (CA) with varying amounts of ethylenediamine (EDA). The degree of functionalization after amide coupling is compared with already reported amine quantification methods. Which revealed a significant difference in accessibility and reactivity for the amide coupling reaction.

Institute and/or researcher Twitter usernames: @Trsgtt @jlu_lama @PoliTOnews